# The economic dependency of the Bitcoin security[1]

Pavel Ciaian

(corresponding author)

European Commission, Joint Research Centre (JRC)

Via E. Fermi 2749, 21027 Ispra, Italy

*e-mail: pavel.ciaian@ec.europa.eu*

d'Artis Kancs

European Commission, Joint Research Centre (JRC)

Via E. Fermi 2749, 21027 Ispra, Italy

*e-mail: d'artis.kancs@ec.europa.eu*

Miroslava Rajcaniova

Slovak University of Agriculture in Nitra (SUA) and University of West Bohemia (UWB)

Tr. Andreja Hlinku 2, 949 76 Nitra, Slovakia

*e-mail: miroslava.rajcaniova@uniag.sk*

# The economic dependency of the Bitcoin security

**Abstract:**

We study to what extent the Bitcoin blockchain security permanently depends on the underlying distribution of cryptocurrency market outcomes. We use daily blockchain and Bitcoin data for 2014-2019 and employ the ARDL approach. We test three equilibrium hypotheses: (i) sensitivity of the Bitcoin blockchain to mining reward; (ii) security outcomes of the Bitcoin blockchain and the proof-of-work cost; and (iii) the speed of adjustment of the Bitcoin blockchain security to deviations from the equilibrium path. Our results suggest that the Bitcoin price and mining rewards are intrinsically linked to Bitcoin security outcomes. The Bitcoin blockchain security's dependency on mining costs is geographically differenced – it is more significant for the global mining leader China than for other world regions. After input or output price shocks, the Bitcoin blockchain security reverts to its equilibrium security level.

**Keywords:** Bitcoin, blockchain, proof-of-work, ARDL, institutional governance technology

**JEL classification:** D82, G12, G15, G29

**Introduction**

The Bitcoin blockchain is a distributed alternative to centralized transaction-recording and record-keeping systems by enabling trustworthy interactions, recording transactions among non-trusting parties and storing records. The underlying ledger that creates and stores records of transactions is a digital chain of blocks, where information is recorded sequentially in data structures known as 'blocks' stored into a public database ('chain'). Being distributed, blockchain is run by a peer-to-peer network of nodes (computers) who collectively adhere to an agreed distributed validation algorithm (protocol) to ensure the validity of transactions. Given that a distributed network of anonymous record-keeping peers (miners) with free entry and exit is inherently 'trustless', it requires some trust-enhancing mechanism. The trust problem among non-trusting parties is solved by requiring miners to pay a cost (in form of computing power for blockchain) to record transaction information and requiring that future record-keepers (miners) validate those reports. Under a well-functioning institutional governance technology, blockchain is immutable, meaning that once data have been recorded on the blockchain, they cannot be altered anymore.[2]

Ensuring a transaction correctness and security, enforcing property rights and contracts are preconditions for a functioning of markets. In traditional centralized institutional governance systems, typically, state or other centralized intermediary guarantees the transfers of ownership ensures transfers of possession and guarantees the security of property rights and contract enforcement. The correctness and security is incentivized via monopoly rents. A comparative advantage of distributed institutional governance systems such as blockchain is the ability to achieve and enforce a uniform view (agreement) among non-trusting parties with divergent interests and incentives on the state of transactions in a cost-efficient and consensus-effective way. Blockchain security algorithms make it possible for distributed record-keepers to confirm that the network rules are being followed, i.e. all other record-keepers ignore any chain containing a block that does not conform to the network rules. The correctness and security is incentivized via physical resource costs – the proof-of-work (PoW) makes it costly to extend invalid chains of blocks (Davidson, De Filippi and Potts 2016; Cong and He 2018; Derks et al. 2018).[3]

---
[2] For more conceptual discussion see Appendix A1

In the same time, ensuring a transaction correctness and security may be more challenging for distributed digital ledgers than for traditional centralized ledgers (Abadi and Brunnermeier 2018).[4] First, because digital goods have two characteristics – non-rivalry and non-excludability – which compared to traditional private goods do not prevent a double spending. Second, the security budget of distributed ledgers is endogenous and fluctuates over time (in a fiat currency nomination), implying that the underlying institutional governance technology may become vulnerable to attacks in a low-price and low-security-budget equilibrium. Hence, ensuring the transaction correctness and security may become an issue particularly in periods of low security budget. Indeed, a number of cryptocurrency-blockchains with a relatively small security budget of preventing attacks have experienced successful majority (hash rate)[5] attacks in recent years, e.g. Bitcoin Gold, Ethereum Classic.[6]

---

[3] There are two prominent designs for validation mechanism – proof-of-work (PoW) and proof-of stake (PoS) – with each having different incentive scheme in achieving consensus. In this paper we focus on the PoW linked to Bitcoin which is the largest and most popular cryptocurrency.

[4] In the context of creating and maintaining distributed ledgers of information, a strong security implies immutable records of transactions, including ownership rights and smart contracts.

[5] The hash rate measures the speed at which a given mining machine operates. Usually, the hash rate is expressed in hashes per second (h/s). For example, a mining machine operating at a speed of 100 hashes per second makes 100 guesses per second. Thus, the hash rate measures how much computer power a cryptocurrency network is deploying to continuously solve the computational problem and generate/record blocks. For example, currently the Bitcoin hash rate is around 110 million Tera per second where 1 Tera/s is equal to one trillion hashes per second.

[6] For example, Bitcoin Gold, a hard fork of Bitcoin, experienced a sequence of double-spending attacks in May 2018. Its price measured in USD at the end of that month was 40% lower. Ethereum Classic also experienced a double-spend attack and several deep block reorganizations, following a 50% decline in its price and hash rate in January 2019.

Our study contributes to the literature that has studied PoW blockchain security concerns from a crypto-coin user perspective (see Lee 2019, for a survey). It has been found that crypto-coin users value security and internalize and price the risk of a blockchain attack that could compromise the ability to exchange crypto-coins for goods. Blockchain users who engage in on-chain transactions value security measured by the amount of computational power committed to the blockchain; ceteris paribus they prefer more computing power being committed to the ledger. However, there is little empirical evidence available in this literature about the economic dependency of the blockchain security (see Abadi and Brunnermeier 2018; Iyidogan 2020; Pagnotta 2020, for theoretical analyses). Moreover, there is confusion in this literature that the blockchain security would be an embedded property of the underlying institutional governance's technology.

To close this research gap, the present study investigates the economic dependency of the Bitcoin blockchain security. To what extent the digital ledger's record-keeping security budgets (measured by mining rewards in a fiat currency nomination) of Bitcoin is correlated with the cryptocurrency market outcomes? We estimate empirically the extent to which this relationship is contingent upon economic incentives by using daily Bitcoin data for 2014-2019 and employ an autoregressive distributed lag approach that allows treating all the relevant moments of the blockchain series as potentially endogenous.

The paper is organized as follows. The next section presents the testable hypothesis. The third section presents econometric approach followed by data description. The fifth section presents the estimation results, while the finial section concludes.

**Conceptual Framework: Testable hypotheses**

The Bitcoin blockchain mining consists of record-keepers (called miners) of a distributed network competing for the right to record information about new transactions (in intervals of around ten minutes) to the digital ledger. Miners have to solve a computationally challenging problem in order to record information and validate others' reports. Solving the computational problem (puzzle) is energy intensive and costly. First, miners have to invest in a computer power, causing fixed costs. Mining involves also variable costs, such as energy (and time) for the computationally-intense mining process, and a building rent for the location of the mining equipment. On the revenue side, mining incentives are ensured via rewards for a correct and secure record keeping. The reward for every block is allocated to the miner that first solves the computational problem (hash function), by using guess and check algorithms based on the new and previous blocks of transactions.[7]

The probability of a miner winning the block's mining contest (i.e. the right to record a new block on the ledger and collect the mining reward) depends on a miner's computer power devoted for each block relative to the computer power of other miners. Following the mining models of Thum (2018), Budish (2018) and Ciaian et al. (2021), the total equilibrium computer power devoted to Bitcoin mining can be expressed as:[8]

(1) $$n_t m_t = \left(\frac{1}{n_t}\right)^{\frac{1+\gamma}{1-\gamma}} \left[\gamma(n_t - 1) \frac{E(p_t) R_t}{c_t + (\rho + \delta) q_t}\right]^{\frac{1}{1-\gamma}}$$

where $m_t$ is computer power per miner (e.g. expressed by the number of computer operations), $n_t$ is the total number of miners, $n_t m_t$ is the total computer power devoted to the Bitcoin mining, $E(p_t)$ is the expected Bitcoin price, $R_t$ is mining reward (Bitcoin quantity), $c_t$ is variable costs per computer operation (e.g. energy cost), $\delta$ is depreciation rate, $\rho$ is a discount rate for time preference, $\gamma$ is a transformation parameter (with $0 < \gamma \leq 0$) determining the odds of winning a block between big and small miners, and $q_t$ is purchase price of one unit of a computer equipment of a given efficiency, $\varepsilon$.

---

[7] For more conceptual discussion see Appendix A1.
[8] For derivations in Appendix A2.

Equation (1) implies that the total computer power devoted to the Bitcoin mining increases in the relative gain from mining, $E(p_t)R_t/(c_t + (\rho + \delta)q)$ and that it fluctuates with the Bitcoin price. Ceteris paribus, higher nominal reward or higher Bitcoin price (costs of mining) induces miners to invest in more (less) computing resources. The opposite is true when agents anticipate the value of Bitcoins to be low, miners have lower incentive to invest in computational resources, the competition and network hash rate decline and the security of the network decreases. Equation (1) implies that miners have incentives to revert to the equilibrium level of computer power as a response to Bitcoin price changes because otherwise miners would experience losses. Further, the total computer power increases at a decreasing rate in the level (intensity) of miners' competition, $(n_t - 1)/n_t^2$.

Equation (1) defines the equilibrium behavior of honest miners by pinning down how much computer power they would allocate to mining for a given value of reward and competition from other miners. The total computer power devoted to the blockchain mining, $n_t m_t$, determines the security equilibrium of blockchain. The more challenging is the computational mining puzzle to solve, the safer and more stable is the blockchain's institutional governance technology because it becomes more costly for a potentially dishonest miner to conduct an attack.

A successful attack may adversely affect the perception of Bitcoin by its users reducing the trust and hence valuation of cryptocurrency. If the reduction of the trust is sufficiently large, it may cause a collapse in the economic value (price) of Bitcoin. As equation (1) implies, the amount of computer power for mining and hence the hash rate of the network would reduce, which might eventually lead to a collapse of Bitcoin blockchain. Thus, the security of Bitcoin blockchain is dependent on the size of mining reward received by miners which also determines the total computer power determined in equation (1).

Following these analyses, we can derive three testable hypotheses:

- Hypothesis 1: Security outcomes of the Bitcoin blockchain. If agents anticipate the value of Bitcoin to be low, miners have little incentive to invest in computational resources, and the security of the network is low. The opposite is true when agents anticipate the value of Bitcoin to be high.

  *Ceteris paribus, the blockchain security is sensitive (elastic) to the mining reward.*

- Hypothesis 2: The physical resource cost to write on the Bitcoin blockchain is intrinsically linked to the cost of preventing attacks; the security of blockchain is structurally linked to the ledger's security budget and mining costs.

  *Ceteris paribus, the blockchain security is sensitive (elastic) to mining costs.*

- Hypothesis 3: *Ceteris paribus, the Bitcoin blockchain security adjusts quickly to deviations from the equilibrium.*

**Estimation strategy**

Equation (1) implies that the security (measured by the allocated computer capacity) of the Bitcoin blockchain depends on mining rewards, the intensity of miners' competition, mining costs, discount rate and the computer equipment cost-efficiency. Applying a logarithmic transformation to equation (1), yields the following equilibrium relationship:

(2) $$y_t = b_0 + \beta x_t + u_t$$

where *y* represents the dependent variable – the Bitcoin blockchain security (computer capacity devoted to mining), $\beta$ is a vector of coefficients to be estimated, *x* is a vector of explanatory covariates – mining rewards, $p_t R_t$, the number of miners, $n_t$, the intensity of miners' competition, $(n_t - 1)/n_t^2$, the cost of mining (including the discount rate), $c_t + (\rho + \delta)q_t$ and the commuter equipment efficiency, $\varepsilon_t$, and $u_t$ is an error term.

Equation (2) implies that the coefficients associated with the mining reward and the intensity of miners' competition are expected to be positive (*number of miners* and *mining reward* *effects* in Figure 1). In contrast, the coefficient linked to the cost of mining (energy costs and discount rate) is expected to be negative (*mining cost effect* in Figure 1). The computer equipment cost-efficiency coefficient is expected to have a positive relationship with the computer power, because everything else constant, higher computing efficiency implies that less energy is needed to achieve a certain computing hash rate.

Our primary interest is the coefficient associated with the mining reward and the cost of mining: the former measures the elasticity of the Bitcoin blockchain security (mining network hash rate) with respect to the mining reward (Hypothesis 1), whereas the later with the mining costs (Hypothesis 2). They reflect the level of dependency of the security of the Bitcoin blockchain with respect to Bitcoin market outcomes.

*Estimation issues*

The estimation of the economic dependency of the blockchain security described by equation (2) is subject to several econometric issues. Our first concern is the endogeneity problem. The endogeneity issue is particularly relevant for our data series, as the security outcomes of the Bitcoin blockchain and the mining reward may be determined simultaneously. For example, if agents anticipate the value of Bitcoin to be low, miners have little incentive to invest in computational resources, and the security of the blockchain would be low. In that case, crypto-coin users may not wish to accumulate large real balances, and the resulting market valuation for Bitcoin would be low. The opposite would be true when agents anticipate the value of Bitcoin to be high.

In order to address this endogeneity concern, we employ the Autoregressive Distributed Lag (ARDL) methodology that is being increasingly used for studying financial markets (e.g. Stoian and Iorgulescu 2020). We employ the ARDL bounds testing approach developed by Pesaran et al. (2001) to estimate the blockchain security equilibrium relationship (2) as it enables to model the long- and short-run relationships simultaneously and has several advantages over the standard cointegration methods. A key advantage for our analysis is that the ARDL approach allows treating all the relevant moments of blockchain series as potentially endogenous.[9]

In the context of cryptocurrencies, another important advantage of the ARDL approach is that it permits different numbers of lags for each data series. Contrary to other comparable cointegration methods, the ARDL methodology does not require testing for the order of integration; it can be applied irrespective of whether the regressors are purely I(0), purely I(1) or mutually cointegrated variables (Pesaran et al., 2001). However, Ouattara (2004) argues that if I(2) variables are present in the data, the computed $F$ statistics of Pesaran et al. (2001) become invalid. To make sure that none of the variables is integrated of order I(2) or beyond, we test the stationarity of series and their first differences using the augmented Dickey–Fuller (ADF) test, the Dickey–Fuller GLS test (DF-GLS) and Phillips–Perron (PP) test. The appropriate number of lags for the series is determined by the Akaike Information Criterion. Accordingly, the role of the Bitcoin mining reward and the proof-of-work cost for each of the respective moments can be estimated after accounting for the information embedded in the lags of the entire distribution of blockchain security outcomes.

Another concern is a potential errors-in-variables problem because part of the series is obtained from primary non-harmonized data sources and it is difficult to judge how reliable these series are. In particular, this concerns those series that are not recorded on blockchain, such as, mining cost data. Indeed, the time series measuring variable mining unit costs in different world regions are collected by using different sampling methodologies and different weights. Some of these issues can be overcome by first differencing the data. Nevertheless, part of potential errors-in-variables issues remain. To address the remaining errors-in-variables issues, we construct alternative proxies for measuring the dependent variable – blockchain security – and key explanatory variables – mining competition – and estimate these otherwise identical mining models for robustness.

---

[9] As noted by Pesaran and Shin (1999, p. 16), the use of ARDL is well suitable to address the endogeneity problem: *"appropriate modification of the orders of the ARDL model is sufficient to simultaneously correct for residual serial correlation and the problem of endogenous regressors"*.

*Econometric strategy*

The ARDL bounds testing procedure is applied to test the existence of a long-run relationship. The general form of an ARDL(*g, z,.....,z*) model is standard and follows the literature:

(3) $$y_t = b_0 + \sum_{i=1}^{g} \phi y_{t-1} + \sum_{i=0}^{z} \beta_i x_{t-1} + u_t$$

where *y* represents the dependent variable – security (computer power) of mining, *x* is a vector of independent variables – mining rewards, intensity of miners' competition, energy costs, discount rate and the commuter equipment efficiency, *g* is the number of optimal lags of the dependent variable and *z* represent the number of optimal lags of each explanatory variable.

Pesaran et al. (2001) has proposed two types of critical values for a given significance level. The first type assumes that all variables in the model are I(1), whereas the other assuming that all series are I(0). If the computed *F* statistic is below the lower bound, the null hypothesis of no long-run relationship fails to be rejected. In such case, an ARDL model in first differences without an error correction term should be estimated. If the *F*-statistic lies between the two bounds, the result is inconclusive. If the computed *F*-statistic exceeds the upper bound, the null hypothesis of no cointegration is rejected. In this case, the error correction model to be estimated is:

(4) $$\Delta y_t = b_0 - \alpha(y_{t-1} - \theta x_t) + \sum_{i=1}^{g-1} \psi_{yi} \Delta y_{t-i} + \sum_{i=0}^{z-1} \psi_{xi} \Delta x_{t-i} + u_t$$

where $\theta$ represent the long-run coefficients, $\psi$ are short-run multipliers and $\alpha$ shows the speed of adjustment of the dependent variable to a short-term shock. It measures how quickly the blockchain security adjusts to deviations from the equilibrium (Hypothesis 3).

Following the standard procedure in the literature (Pesaran et al. 2001), we apply a set of diagnostic tests, as the validity of ARDL results is based on the assumption of normally distributed error terms, no serial correlation, heteroscedasticity and stability of the coefficients. The empirically estimable specification of the models and the number of lags is determined in accordance with the results of diagnostic tests, including Breusch-Godfrey LM test and Durbin's alternative test for autocorrelation, Breusch-Pagan/Cook-Weisberg test for heteroscedasticity, normality testing and cumulative sum test for the parameter stability.

**Data**

In empirical estimations, we use daily data for the period 27/12/2014 – 4/9/2019. The details of data series used in estimations and their sources are reported in Table 1. All time-series are transformed in a log-form in the estimations, implying that the estimated coefficients can be interpreted as elasticities. Table 2 provides a descriptive statistic of the data used.

Our main dependent variable measuring the computer power devoted for mining is *hash rate* and is represented in average daily hashes per second.[10] For robustness, we also consider the mining *difficulty* as an alternative dependent variable which measures the effort required to mine a new block for the blockchain.

Following equation (2), our independent variables include the *number of miners* and the derived *competition intensity*, $(n_t - 1)/n_t^2$. We also consider alternative proxies for competition intensity – Herfindahl-Hirschman index (*hhi*) and normalized Herfindahl-

---

[10] Hash rate measures the speed at which mining machines operates. Usually, the hash rate is expressed in hashes per second (h/s). For example, a mining machine operating at a speed of 100 hashes per second makes 100 guesses per second.[10] Thus, the hash rate measures how much computer power the Bitcoin network is deploying to continuously solve the computational problem and generate/record blocks.

Hirschman index (*hhi normalised*) – in order to account for unequal distribution of computer power between different miners. The variable *mining reward* is measured as the average daily value of the reward per block calculated by dividing the total mining reward per day (in US dollars) by the total number of blocks per day[11]. To measure a region-specific cost of mining, we use electricity prices in Europe (*electricity Europe*), China (*electricity China*) and North America (*electricity N. America*). We proxy discount rate with the US 10-year treasury constant maturity rate (*10-year-treasury*). Finally, the mining *equipment efficiency* is proxied with the most efficient mining hardware available in each time period measured by the energy efficiency of the hardware (Zadé and Myklebost 2018; CBEI 2019; Delgado-Mohatar, Felis-Rota and Fernández-Herraiz 2019). All variables are used in a log-form in the estimations, implying that the estimated coefficients can be interpreted as elasticities.

**Results**

We start with checking the stationarity properties of our time series, as the ARDL procedure requires all variables to be either I(0) or I(1). The results of the Augmented Dickey-Fuller test, the Dickey–Fuller GLS test (DF-GLS) and the Phillips–Perron (PP) test indicate that there is no variable integrated of the second order and thus we can apply the ARDL approach.

Table 3 summarizes the three estimated mining models with different specification of explanatory variables and for each of the three models we include two sub-models with alternative measures of the Bitcoin blockchain security, i.e. *hash rate* and *difficulty*. The three estimated mining models differ by the proxy measuring the computer intensity. Model 1 uses *competition intensity* variable, $(n_t - 1)/n_t^2$, as derived in equation (1), whereas models 2 and 3 use the two alternative proxies for competition intensity: the Herfindahl-Hirschman index

---

[11] Mining reward contains the total value of coinbase block rewards and transaction fees paid to miners.

(*hhi*) and normalized Herfindahl-Hirschman index (*hhi normalised*), respectively. The rest of variables are the same across all estimated mining models.

*Long-run economic dependency of the Bitcoin security*

*Mining rewards*. In line with theoretical expectations, the ARDL estimation results confirm a long-run structural relationship between the mining reward and security outcomes of the Bitcoin blockchain (Table 4). This holds for both variables measuring the computer power devoted to mining, *hash rate* and *difficulty*, and across all estimated models. The estimated elasticities corresponding to the *mining reward* variable are in the range between 1.4 and 2.0, suggesting an elastic response in the mining computer capacity to permanent changes in the mining reward. That is, a 1% permanent increase in the mining reward increases the underlying blockchain security by 1.4% to 2.0% in the long-run. Hence, our estimates fail to reject Hypothesis 1: the Bitcoin security is overly sensitive (elastic) to Bitcoin mining reward. A change in the payoff from mining causes a more than proportionate change in the Bitcoin security.

Given that usually mining costs are incurred in standard fiat currencies (e.g. US dollar, Euro), the value of the mining reward fluctuates with the price of Bitcoin[12] which in turn affects mining reward and mining incentives. Thus, if the expected Bitcoin price decreases, lower mining incentives reduce the equilibrium computer mining capacity and hence the Bitcoin security. The reverse is valid in the case of a positive Bitcoin price shock.

*Proof-of-work costs*. As regards the variable mining unit costs and security outcomes of blockchain, the results are more nuanced. As explained above, to measure a region-specific the cost of mining, for the variable construction we have used separate electricity prices for

---

[12] Note that the change in Bitcoin price is the main factor deriving the change in the value of mining reward because according to the algorithm the quantity of mining reward in Bitcoins, $R_t$, changes (halves) only approximately every 4 years, whereas Bitcoin price changes daily.

Europe, China and the North America. The estimation results for the world global mining leader China[13] – 65% of the global Bitcoin hash rate are concentrated in China – imply a statistically significant and positive relationship between the security of the underlying blockchain and the proof-of-work cost (models M2.1 to M3.2). The estimated variable mining unit costs coefficients the North America are statistically significant and negative in models M1.1, M2.1. and M3.1. For Europe, the mining cost coefficients are statistically insignificant. One explanation for these mixed results could be that the intensive margin of mining is larger in China, where all major mining pools are located. A declining number of miners might actually not reduce the mining network hash rate, if those individual miners join the rapidly growing miner pools. Parra-Moyano, Reich and Schmedders (2019) find an empirical evidence for learning by mining, which results in a decreasing extensive margin of mining and an increasing intensive margin of mining. Second, our estimates capture other long-term behavioral effects of miners induced by a permanent change in electricity prices such as shifting mining location to places with cheaper energy (e.g. to remote regions of China, from mainland Europe to Iceland to harvest geothermal power).[14] Such long-term behavioral effects may actually increase computer power, if the energy savings more than offset the price increase. Finally, these results may also reflect the fact that the mining input cost data (which are location-specific) are considerably less reliable than the mining reward data, which are publicly available for every single historical Bitcoin transaction.

Based on these ARDL bounds testing results, we cannot provide a robust answer regarding Hypothesis 2. Instead, these results call for further analysis using more disaggregated location-specific mining cost data, given that the effect of electricity prices in North America is in line with expectations as derived in equation (1), whereas the opposite relationship is found for the electricity prices in China.

---

[13] https://news.bitcoin.com/65-of-global-bitcoin-hashrate-concentrated-in-china/

[14] https://www.vox.com/2019/6/18/18642645/bitcoin-energy-price-renewable-china

*Competition and network externalities.* The variables measuring the miners' competition level have a negative impact on security outcomes of the Bitcoin blockchain; all estimated coefficients are statistically significant. These results suggest that a permanent increase in the competition intensity exercises a downward pressure on the mining computer power in the long-run. As discussed in Appendix A.1.2, digital distributed ledgers such as blockchain are subject to a number of network externalities. When new miners enter the mining of blockchains, two types of direct network externalities related to the blockchain security arise, one positive and one negative. The positive network externality implies that the blockchain security is increasing with the number of miners, because each additional node reinforces the chain's security. In line with the previous literature (Waelbroeck, 2018), the negative network externality occurs because each individual miner invests in the mining-computing power, which increases both the individual miner's marginal income though also mining costs, as the difficulty of the computational problem increases in the number of miners and their computing power ("hash-power"). Increasing the difficulty of mining reduces the incentives for mining and – in the presence of learning by mining – increases the concentration of mining activities, reducing in such a way the blockchain security. Our estimates suggest that the negative network externality dominates of the positive network externality. Our results are in line with those of Parra-Moyano, Reich and Schmedders (2019) who find that the probability of winning a mining contest increases with the miner size. This motives miners to join mining pools to increase their probability to win the mining contest and receive reward.[15] Indeed, our competition proxy variables are constructed based on the observed number of miners but not on the number of members within mining pools. And since a greater competition may imply fewer miners (because many individual miners join mining pools), the implied actual long-run relationship between the competition intensity and mining power becomes negative.

---

[15] Other benefit of joining mining pools is that it creates a steady stream of income, rather than greater income but at lower frequency (i.e. due to lower odd of winning the mining contest) with individual mining (Liu and Wang 2017).

*Hardware efficiency*. In line with the theoretical model, the hardware efficiency variable has a statistically significant negative impact on the mining computer power in models M1.1, M1.2 and M3.1. The estimation results imply that an increase in the efficiency of mining equipment (decrease in the input units of the mining *hardware efficiency* per security output unit) leads to an increase of security outcomes of the Bitcoin blockchain in the long-run. The estimated elasticities vary between -0.6 and -0.8, implying that a 1% permanent increase in the efficiency of mining equipment increases the mining computer power in the long-run by between 0.6% and 0.8%. Hence, the Bitcoin blockchain mining security is dependent of the mining technology available at a given point of time.

*Time preference*. The *10-year-treasury* variable, which is a proxy for the discount rate, has a statistically significant positive impact on the mining computer power. This result could be explained by the fact that the *10-year-treasury* actually captures a miner investment competition effect (i.e. an alternative financial asset return). As far as Bitcoin is perceived as an investment asset, shocks to competing financial asset returns (including *10-year-treasury*) are expected to impact positively miners' choices to invest in the mining of Bitcoin. Our results confirm that Bitcoin is perceived by miners to be competing for investment with other financial assets and thus need to deliver a competitive return. The return arbitrage among alternative investment opportunities implies a positive price relationship between Bitcoin and alternative financial assets (Murphy 2011; Ciaian, Rajcaniova and Kancs 2018). Thus, the positive coefficient estimated for the *10-year-treasury* variable implies that miners are motivated to invest in more computer power for mining when the returns to financial assets increase.

The estimates of the error correction term – which measures the speed of adjustment of the short-run dynamics of mining to the long-run equilibrium path – are statistically significant across all models. The error correction terms vary between -0.002 and -0.009. This means that between 0.2% and 0.9% of the long-run disequilibrium in mining power is corrected by the short-run adjustment the same day. Or the disequilibrium corrects at a speed of convergence of between 0.2% and 0.9% per day. In terms of the duration, any deviation from the long-run equilibrium is corrected in around 111 to 453 days (or in 0.30 to 1.24 years). These results provide support for Hypothesis 3 that security outcomes of the Bitcoin blockchain adjust to deviations from the long-run security equilibrium.

*Short-run economic dependency of the Bitcoin security*

Generally, short-run results are less significant across the estimated ARDL models than long-run results (Table 5). Contrary to our expectations, there is some support that the mining reward affects negatively the mining computer power in the short-run. However, the estimated elasticity is rather small. A 1% positive shock in the Bitcoin mining reward (the third lag) decreases the mining computer power in the short-run by between 0.014% and 0.020%. This negative relationship between the mining reward and mining power could be a result of other effects such as switching mining to other cryptocurrencies (e.g. to Bitcoin cash) when the relative price of Bitcoin to cryptocurrencies decreases.[16] These findings also tend to support the Hypothesis 1: although an inverse relationship is found, the security outcomes of the Bitcoin blockchain shows sensitivity to Bitcoin mining reward even in the short-run. This short-run inverse relationship could be caused by secondary spiral effects induced by Bitcoin price changes – decrease (increase) – as suggested by Kroll, Davey and Felten (2013), through the subsequent loss (gain) of confidence (thrust) in Bitcoin when Bitcoin mining power decreases (increases) which might further reduce (increase) the Bitcoin price.

---

[16] There is some evidence of asymmetric change in Bitcoin and altcoin prices: shocks to altcoins prices tend to be greater than Bitcoin price shocks (Reiff 2018; Cheikh, Zaied and Chevallier 2020). This implies that the relative prices of Bitcoin to altcoins are inversely related with the Bitcoin price changes which may incentivize miners to shift some Bitcoin computer power to mining altcoins when Bitcoin price increase, and shift back the computer power to Bitcoin mining when Bitcoin price declines. Note that the shift in mining between different cryptos is less relevant for ASIC mining hardware, commonly used for Bitcoin mining, which is more efficient in mining specific cryptocurrencies (specific cryptographic hash algorithm) and cannot be used for mining other types of cryptocurrencies.

As expected, the variables measuring mining competition level (*number of miners*, *competition intensity*) have a statistically positive impact on the computer power in models M1.1, M1.2, M.2.2 and M3.2. These results suggest that in the short-run, the competition among miners stimulates deployment of more mining equipment in line with the model derived in equation (1). These results are in contrast to the long-run estimates. While in the short-run the miners' competition leads to expansion of the Bitcoin mining power, in the long-run the inverse relationship is valid indirectly suggesting reduced competition level as individual miners have the incentive to join mining pools.

In line with the theoretical expectation in equation (1) the electricity prices in China have statistically negative impact on the computer power outcomes of the Bitcoin blockchain in the short-run in models M1.1, M2.1 and M3.1. The reverse relationship was estimated in the long-run, where permanent increase in the electricity prices in China led to an increase in the mining power suggesting that other structural changes in miners' behavior might take place when the cost changes are permanent. The estimated variable mining electricity costs for North America and Europe are statistically insignificant in the short-run. These findings support the Hypothesis 2 that the security outcomes of the Bitcoin blockchain is sensitive to mining costs.

---

[16] https://www.vox.com/2019/6/18/18642645/bitcoin-energy-price-renewable-china

The lagged dependent variables (*hash rat*e and *difficulty*) are statistically significant across all models. Their coefficients vary between -0.04 and -0.46. The relatively high values of these coefficients indicate that the short-run shocks in the mining computer power disappear over time relatively fast: in around 2.2 to 28.6 days. These results support the Hypothesis 3 that the security outcomes of the Bitcoin is sensitive to Bitcoin market outcomes in the short-run to fluctuations with instant shocks disappearing over a short time period (within few days).

The short-run effects of *electricity prices*, *hardware efficiency*, *10-year-treasury* and alternative proxies for competition intensity cannot be examined due to the ARDL specifications, as no lags of these dependent variables entered the model.

**Discussion and concluding remarks**

Alongside being a new innovative information and computation technology, blockchain also introduces a new institutional governance technology that makes possible the enforcement of contracts, ownership rights and development of distributed autonomous organizations. Bitcoin blockchain offers an example where the institutional governance system (i.e. the security of the enforcement of property rights and contracts) is endogenously determined by its underlying transaction validation algorithm. This is in a sharp contrast to the traditional centralized institutional governance systems which face many impediments to bring about dynamic institutional changes. While an institutional innovation (creation of institutions) is desirable, the institutional deterioration (destruction of institutions) is not. Sustaining economic transactions built on the Bitcoin blockchain requires to have in place a correct and secure system of verification and enforcement of property rights and contracts. As a result, the security and correctness of the verification and enforcement system should portray a certain neutrality from the fluctuation of economic market outcomes executed on the Bitcoin blockchain in order to reduce a potential risk of adverse attacks and a subsequent loss of trust among blockchain users.

The present paper studies the economic dependency of the Bitcoin security of the enforcement quality on the underlying economic incentives defined within its validation algorithm. We apply time-series analytical mechanisms using data for the period 27/12/2014 – 4/9/2019 to estimate the responsiveness of Bitcoin security to the economic incentives both in the short- and long-run in order to provide empirical understanding for three tested hypothesis: the Bitcoin security is dependent on the mining reward (Hypothesis 1), security outcomes of the Bitcoin and the proof-of-work cost (Hypothesis 2) and any disequilibria in Bitcoin security revert back to its equilibrium relatively fast (Hypothesis 3). We employ an autoregressive distributed lag approach that allows treating all the relevant moments of the blockchain series as potentially endogenous.

Our results suggest that the Bitcoin price and mining rewards are intrinsically linked to blockchain security outcomes. Results for mining costs are geographically differenced – they are more significant for the global mining leader China than for other world regions. The estimates for the speed of adjustment of the Bitcoin security suggest that any disequilibria revert back to its equilibrium relatively fast. Based on the ARDL bounds testing results, we fail to reject Hypothesis 1 and 3, suggesting that Bitcoin security is highly responsive to permanent shocks in mining reward and its adjustment to the permanent and short-term shocks is relatively fast. In the long-run, we find an elastic (between 1.4 and 2.0) and positive response in Bitcoin mining computer power to permanent change in mining reward, while in the short-run, the elasticity is smaller and the relationship is negative (between -0.014% and -0.020%) (Hypothesis 1). Further, any deviation in Bitcoin mining computer power from the long-run equilibrium is corrected in around 0.30 to 1.24 years, whereas the short-run shocks disappear in around 2.2 to 28.6 days (Hypothesis 3). Regarding the Hypothesis 2, we find some support that the mining costs (i.e. electricity prices) impact the Bitcoin mining power and thus its security but the results are not robust across estimated models and the sign of estimated effects is not always consistent with theoretical expectations. This could be explained by the fact that the permanent shocks occurring to the electricity prices might have been accommodated by miners through other behavioral changes (e.g. relocation of mining to locations with less expensive electricity). Further, our estimates show that the Bitcoin security is dependent on the competition intensity among miners and the efficiency of the mining technology. The competition intensity among miners leads to shot-run increase in mining computer power, whereas the reverse is valid in the long-run likely due to the structural changes taking place in mining whereby individual mining is replaced in favor of concentration of computer power to a relatively small number of mining pools. As regards the mining technology, its improvement is found to stimulate the expansion in mining computer power and thus can undermine the Bitcoin security in future if, e.g., a strong innovation would take place in the computing technology (e.g. Quantum Computers).

The findings of this paper challenge the entire Bitcoin security model. Our results suggest that the Bitcoin blockchain security is highly sensitive to the reward system which incentivizes distributed network participants to provide validation and enforcement services by making available their computer capacity. As the ARDL estimates show, these services are highly responsive and adjust fast to the internal Bitcoin economics (i.e. to changes in the mining reward). This may pose problems particularly in low-security equilibriums when the mining reward value declines. In such a situation, the incentives for supplying security services descrease and may make the transactions executed on Bitcoin blockchain vulnerable to potential attacks. The Bitcoin security is sustainable only if the value of reward increases over time particularly given that by design of the Bitcoin algorithm, the reward is halved approximately every 4 years (every 210,000 blocks). This is reinforced by the fact that the innovations in the computing technology may significantly reduce the cost of supplying computer power and thus might also undermine Bitcoin strength against potential hostile attacks. Hence, in the medium-run, the entire security concept of the Bitcoin may need to be redesigned in order to reduce as much as possible its vulnerability to proof-of-work costs and cryptocurrency market outcomes. Otherwise, the Bitcoin blockchain might fail to generate a sufficient trust to provide incentives that would attract economic agents to develop activities on the distributed ledger and thus to stimulate the growth of the distributed digital economy.

**Acknowledgement**

We gratefully acknowledge the financial support received from the Scientific Grant Agency under the grant VEGA 1/0422/19 (Slovak Republic) and GAČR project 19-18080S (Czechia).

**Table 1. Data sources**

| Variable | Unit | Name of variable | Source | Description |
|---|---|---|---|---|
| *Dependent variables* | | | | |
| hash rate | Hash/second | Total computer power | bitinfocharts.com | |
| difficulty | Average difficulty per day | Mining difficulty | bitinfocharts.com | |
| *Explanatory variables* | | | | |
| number of miners | No | Number of miners, $n_t$ | blockchair.com | Number of miners, total |
| competition intensity | Index | Competition intensity | Calculated: $(n_t - 1)/n_t^2$ | Calculated based on n_miners_t (n_miners_t – 1)/ (n_miners_t*n_miners_t) |
| hhi | Index | Herfindahl-Hirschman index | Calculated based on $n_t$ and hashrate | |
| hhi normalised | Index | Normalised Herfindahl-Hirschman index | Calculated based on on $n_t$ and hashrate | |
| mining reward | USD per block | Mining reward per block | blockchair.com | Total mining reward per day (in USD) divided by the total number of blocks per day *reward_usd/ no_bl_total* |
| electricity Europe | EUR/MWh | European Electricity Index | www.epexspot.com | |
| electricity China | USD/kWh | Chengdu's Usage Price Electricity for Industry, USD | www.ceicdata.com | |
| electricity N. America | CAD/MWh | Electricity price in North America | reports.ieso.ca | |
| 10-year-treasury | % | 10-Year Treasury Constant Maturity Rate (DGS10) | fred.stlouisfed.org | 10-Year Treasury Constant Maturity Rate (DGS10) |
| hardware efficiency | J/Giga hash | Mining equipment efficiency | Constructed based on: Zadé and Myklebost (2018), CBEI (2019) and Delgado-Mohatar, Felis-Rota and Fernández-Herraiz (2019) | Bitcoin mining hardware generation (most efficient device in each time period), J/Giga hash |

**Table 2. Descriptive statistics of used data**

| Variable | Obs | Mean | Std. Dev. | Min | Max |
|---|---|---|---|---|---|
| hashrate | 3,174 | 38.313 | 5.995 | 25.442 | 45.905 |
| difficulty | 3,174 | 22.456 | 6.020 | 9.581 | 30.008 |
| number of miners | 3,174 | 2.691 | 0.881 | 0.000 | 3.526 |
| competition intensity | 3,174 | -3.144 | 1.018 | -6.908 | -1.386 |
| hhi | 3,174 | -1.638 | 0.720 | -2.608 | 0.000 |
| hhi normalised | 3,174 | -1.958 | 0.860 | -3.176 | 0.000 |
| mining_reward | 3,174 | 8.732 | 2.252 | -0.692 | 12.794 |
| electricity Europe | 3,174 | 3.505 | 0.876 | -6.908 | 4.891 |
| electricity China | 3,174 | -2.179 | 0.063 | -2.402 | -2.113 |
| electricity N. America | 3,174 | 2.554 | 1.944 | -6.908 | 5.617 |
| 10-year-treasury | 3,174 | 0.827 | 0.208 | 0.315 | 1.322 |
| hardware efficiency | 3,174 | 0.151 | 2.943 | -3.219 | 6.240 |

**Table 3. Specification of the estimated models**

|  | M1.1 | M1.2 | M2.1 | M2.2 | M3.1 | M3.2 |
|---|---|---|---|---|---|---|
| hashrate | X |  | X |  | X |  |
| difficulty |  | X |  | X |  | X |
| *Explanatory variables* |  |  |  |  |  |  |
| number of miners | X | X | X | X | X | X |
| competition intensity | X | X |  |  |  |  |
| hhi |  |  | X | X |  |  |
| hhi normalised |  |  |  |  | X | X |
| mining reward | X | X | X | X | X | X |
| electricity Europe | X | X | X | X | X | X |
| electricity China | X | X | X | X | X | X |
| electricity N. America | X | X | X | X | X | X |
| 10-year-treasury | X | X | X | X | X | X |
| hardware efficiency | X | X | X | X | X | X |

**Table 4. Estimation results: long-run impacts**

|  | M1.1 | M1.2 | M2.1 | M2.2 | M3.1 | M3.2 |
|---|---|---|---|---|---|---|
| number of miners | -0.154 | -0.108 | -2.911*** | -3.910*** | -2.415*** | -3.462*** |
| competition intensity | -1.203*** | -1.197*** |  |  |  |  |
| hhi |  |  | -3.707** | -5.996*** |  |  |
| hhi normalised |  |  |  |  | -2.285** | -4.376*** |
| mining reward | 1.398*** | 1.670*** | 1.373*** | 1.988*** | 1.379*** | 1.988*** |
| electricity Europe | -0.234 | 0.136 | -0.338 | 0.161 | -0.337 | 0.185 |
| electricity China | 3.643 | 4.415 | 11.095** | 12.605** | 13.874** | 19.150*** |
| electricity N. America | -0.135* | -0.060 | -0.218** | -0.118 | -0.235** | -0.145 |
| 10-year-treasury | 2.204** | 2.691*** | 5.125*** | 5.457*** | 5.763*** | 7.081*** |
| Hardware efficiency | -0.799*** | -0.588*** | -0.563 | 0.138 | -0.663* | 0.080 |
| *Error correction term* |  |  |  |  |  |  |
| hash rate (-1) | -0.009** |  | -0.007** |  | -0.007** |  |
| difficulty (-1) |  | -0.003** |  | -0.002** |  | -0.002** |
| Speed-of-adjustment (days) | 111 | 333 | 147 | 407 | 151 | 453 |

Dependent variables are hash rate (M1.1, M2.1, M3.1) or difficulty (M1.2, M2.2, M3.2); Speed-of-adjustment is calculated based on error correction rate.

***significant at 1% level, **significant at 5% level, *significant at 10% level. Empty cell indicates absence of a variable in the respective model.

**Table 5. Estimation results: short-run impacts**

|  | M1.1 | M1.2 | M2.1 | M2.2 | M3.1 | M3.2 |
|---|---|---|---|---|---|---|
| Δ hash rate (-1) | -0.440*** |  | -0.428** |  | -0.429** |  |
| Δ hash rate (-2) | -0.455*** |  | -0.443** |  | -0.444** |  |
| Δ hash rate (-3) | -0.320*** |  | -0.305** |  | -0.306** |  |
| Δ hash rate (-4) | -0.221*** |  | -0.206** |  | -0.205** |  |
| Δ hash rate (-5) | -0.166*** |  | -0.151** |  | -0.150** |  |
| Δ hash rate (-6) | -0.104*** |  | -0.091** |  | -0.090** |  |
| Δ hash rate (-7) | -0.075*** |  | -0.064** |  | -0.066** |  |
| Δ difficulty (-1) |  | 0.200** |  | 0.212*** |  | 0.213** |
| Δ difficulty (-2) |  | -0.142** |  | -0.142*** |  | -0.142** |
| Δ difficulty (-3) |  | -0.040* |  | -0.036** |  | -0.035 |
| Δ difficulty (-4) |  | -0.073** |  | -0.072*** |  | -0.071** |
| Δ difficulty (-5) |  | -0.056** |  | -0.055*** |  | -0.054** |
| Δ difficulty (-6) |  | -0.052** |  | -0.049*** |  | -0.049** |
| Δ difficulty (-7) |  | -0.059** |  | -0.055*** |  | -0.054** |
| Δ number of miners |  |  |  | -0.002 |  | -0.003 |
| Δ number of miners (-1) |  |  |  | 0.075*** |  | 0.074** |
| Δ number of miners (-2) |  |  |  | 0.091*** |  | 0.090** |
| Δ competition intensity | 0.010 | -0.001 |  |  |  |  |
| Δ competition intensity (-1) | 0.066*** | 0.032** |  |  |  |  |
| Δ competition intensity (-2) | 0.040*** | 0.033** |  |  |  |  |
| Δ mining reward | -0.019*** | -0.001 | -0.017* | 0.000 | -0.017* | 0.000 |
| Δ mining reward (-1) | -0.020*** | -0.003 |  | -0.002 | -0.016* | -0.002 |
| Δ mining reward (-2) |  | 0.000 |  | 0.001 |  | 0.001 |
| Δ mining reward (-3) |  | -0.014** |  | -0.014*** |  | -0.014** |
| Δ mining reward (-4) |  | -0.015** |  | -0.015*** |  | -0.015** |
| Δ electricity China | -1.246** |  | -1.256* |  | -1.243* |  |
| constant | 0.289*** | 0.046 | 0.357** | 0.078*** | 0.387** | 0.099** |

Dependent variables are hash rate (M1.1, M2.1, M3.1) or difficulty (M1.2, M2.2, M3.2);
***significant at 1% level, **significant at 5% level, *significant at 10% level. Empty cell indicates either absence of a variable in the respective model or the coefficient or the variable is not selected in the estimation; Δ is difference.

**Figure 1. Interdependency between Bitcoin price, mining costs and mining security**

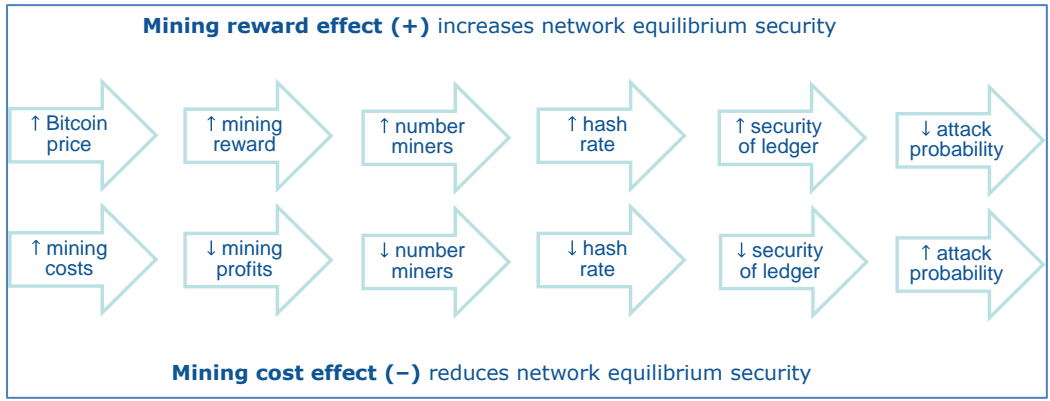

Source: Conceptual framework (section 2 and Appendix. A.1 Model of the Bitcoin mining).

**Figure 2. Dynamics of the Bitcoin price, mining reward, hash rate and network difficulty**

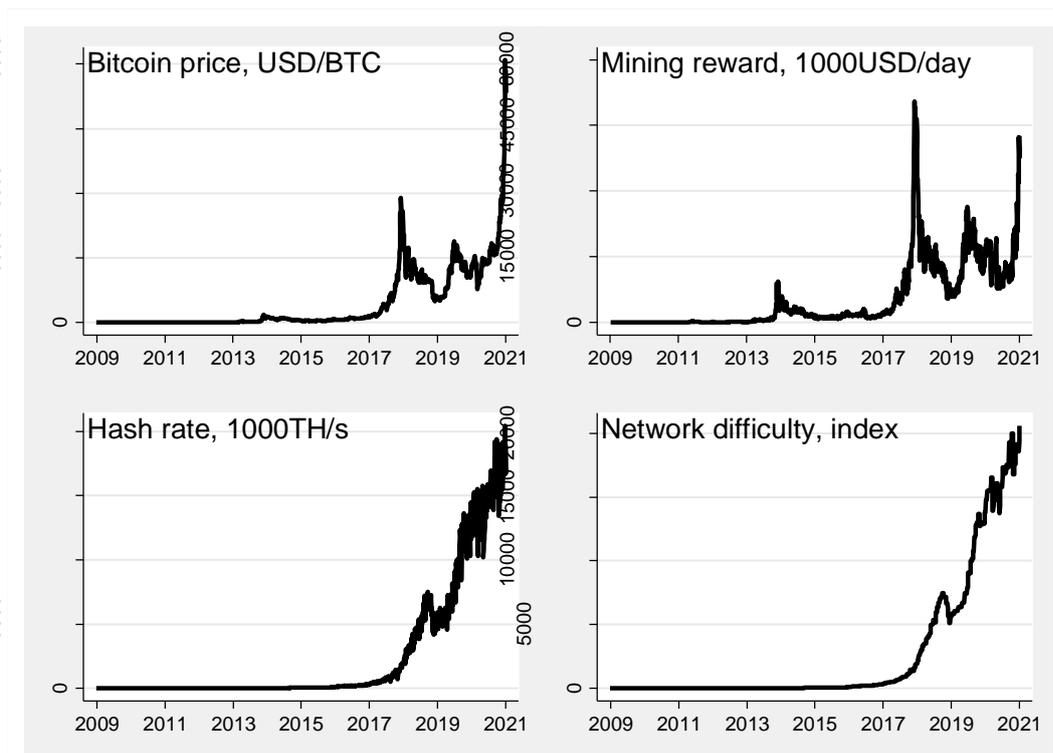

Source: Based on data from blockchain.com/data/.

## Appendix. A.1 Model of the Bitcoin mining

In this Appendix we show a theoretical model to determine equilibrium relationships between Bitcoin blockchain security, market outcomes and resources devoted to the blockchain mining. Building on the mining models of Thum (2018), Budish (2018) and Schilling and Uhlig (2019), we model a rational miner *i* that decides on the quantity of computer power, $m_{it}$ (e.g. expressed by the number of computer operations), to devote for mining each Bitcoin block *t* (represented in block time measured in 10 minute interval which is the average time needed to mine a block in blockchain). The mining output is measured in capacity of blockchain security units.

The probability of miner *i* winning the contest (i.e. the right to generate a new block and collect reward) depends on his/her computer power devoted for each block relative to the computer power of other miners. Previous studies assume that the probability of winning the contest and validating a block is independent of the miner size: $m_{it}/(m_{it}+\Sigma_{j\neq i}^{n_t} m_{jt})$, where $n_t$ is the total number of miners and $\Sigma_{j\neq i}^{n_t} m_{jt}$ is the total computer power of other miners devoted to the block *t* (e.g. Cocco and Marchesi 2016; Thum 2018; Schilling and Uhlig 2019). However, Parra-Moyano, Reich and Schmedders (2019) show that the probability of relatively bigger miners winning the mining contest is higher than that of relatively smaller miners because there is a "learning" effect when mining a particular block with larger mining computers learning faster than smaller mining computers. To account for the *learning by mining*, we assume the following transformation of the probability for a miner winning a block: $e^{m_{it}^{\gamma}}/\left(e^{m_{it}^{\gamma}}+\Sigma_{j\neq i}^{n_t} e^{m_{jt}^{\gamma}}\right)$, where $\gamma$ is a transformation parameter (with $0 < \gamma \leq 0$), which implies that the ratio of odds between big and small miners (mining computers) of winning a block increases with the miners' size, $m_{it}$, while keeping the ratio of miner' size between miners fixed.

The purchase price of one unit of a computer equipment of a given efficiency, $\varepsilon$, is denoted by $q_t$. The successful miner receives reward $p_t R_t$, where $R_t$ is Bitcoin quantity and $p_t$ is Bitcoin price per one unit expressed in monetary values (e.g. US dollar). Miner *i* chooses computer power, $m_{it}$, for a given computer efficiency, $\varepsilon$, so that to maximize the present discounted value of the flow of profits over the infinite time horizon:

(5) $$\pi_i = \Sigma_t \left(\frac{1}{1+\rho}\right)^t \left(\frac{e^{m_{it}^{\gamma}}}{e^{m_{it}^{\gamma}}+\Sigma_{j\neq i}^{n_t} e^{m_{jt}^{\gamma}}} E(p_t)R_t - c_t m_{it} - q_t I_{it}\right) - F$$

Subject to $m_{it+1}$ units of computer power:

(6) $$m_{it+1} = (1-\delta)m_{it} + I_{it}$$

where $c_t$ is variable costs per computer operation (e.g. energy cost), $E(p_t)$ is the expected Bitcoin price, $\delta$ is depreciation rate, $I_{it}$ is investment in computer equipment, $F$ are one-time fixed costs (e.g. building), and $\rho$ is a discount rate for time preference. Deviations from the expected price are random shocks, $v$, with an expected value of zero: $E(p_t) = p_t^*$, where $p_t = p_t^* + v$. We assume a rational price expectation framework of Muth (1961) in which miners base their Bitcoin price formation on all the available information at the time when making their decisions on the investment in $m_i$. Miners are identical, risk-neutral, non-cooperative and profit-driven agents that invest according to the anticipated real value of block rewards.

Maximizing miner $i$'s profits for the given computer power of all other miners yields the following optimal conditions:

$$(7) \qquad -q_t + \frac{1}{1+\rho}\lambda_{t+1} = 0$$

$$(8) \qquad \frac{\gamma m_{it}^{\gamma-1} e^{m_{it}^\gamma} \sum_{j \neq i}^{n_t} e^{m_{jt}^\gamma}}{\left(e^{m_{it}^\gamma} + \sum_{j \neq i}^{n_t} e^{m_{jt}^\gamma}\right)^2} E(p_t) R_t - c_t + \frac{1}{1+\rho}(1-\delta)\lambda_{t+1} = \lambda_t$$

$$(9) \qquad m_{it+1} = (1-\delta)m_{it} + I_{it}$$

where $\lambda_t$ is a shadow price for a unit computer power.

Assuming a steady state equilibrium with $m_{it} = m_{il}$, $R_t = R_l$, $E(p_t) = E(p_l)$, $q_t = q_l$, and $n_t = n_l$ for $t \neq l$ and a symmetric equilibrium with $m_{it} = m_{jl}$, the equilibrium computer power per miner can be derived from equations (7) to (9) as follows:

$$(10) \qquad m_t = \left[\frac{\gamma(n_t-1)}{n_t^2} \frac{E(p_t)R_t}{c_t+(\rho+\delta)q_t}\right]^{\frac{1}{1-\gamma}}$$

Rewriting equation (10) in terms of the total computer power devoted to mining, $n_t m_t^*$, yields the mining equilibrium:

$$(11) \qquad n_t m_t = \left(\frac{1}{n_t}\right)^{\frac{1+\gamma}{1-\gamma}} \left[\gamma(n_t-1)\frac{E(p_t)R_t}{c_t+(\rho+\delta)q_t}\right]^{\frac{1}{1-\gamma}}$$

Equation (11) implies that the total computer power increases in the relative gain from mining, $E(p_t)R_t/(c_t + (\rho + \delta)q)$. The mining equilibrium implies that the computer power devoted to mining fluctuates with the Bitcoin price. This model feature reflects the intuition that, ceteris paribus, higher nominal reward or higher Bitcoin price induces miners to invest in more computing resources. The opposite is true when agents anticipate the value of Bitcoins to be low, miners have little incentive to invest in computational resources, and the security of the network is low.

Second, the mining equilibrium (11) implies that the total computer power increases at a decreasing rate in the level (intensity) of competition, $(n_t - 1)/n_t^2$.

Third, equation (11) implies that miners have incentives to revert to the equilibrium level of computer power as a response to Bitcoin price changes because otherwise miners would experience losses.

We follow Abadi and Brunnermeier (2018) and assume a free entry equilibrium where miners enter until profits are driven to zero. In the blockchain system, miners don't compete in prices but in capacity, similar to Cournot-type firms. An increase in the processing power of competing miners results in the expansion of the total computing capacity. In the presence of network externalities, free entry of miners serves to pin down the strength of the security.

Using equations (9) and (10), it is possible to derive the equilibrium number of miners, $n_t$, depending on mining returns, variable costs, fixed costs and the level (intensity) of competition, $(n_t - 1)/n_t^2$:

$$(12) \quad n_t = E(p_t)R_t / \left( \rho F + (c_t + \delta q_t) \left[ \frac{\gamma(n_t-1)}{n_t^2} \frac{E(p)R}{c_t+(\rho+\delta)q_t} \right]^{\frac{1}{1-\gamma}} \right)$$

Fixed costs are related to credit constraint and rigidities to increase capacity related to financing the entry costs into the mining.

Equations (7) to (11) define the equilibrium behavior of honest miners by pinning down how much computer power they would allocate for mining for a given value of reward and competition from other miners. The total computer power devoted to the blockchain mining, $n_t m_t$, determines the security of blockchain. As discussed above, the more challenging is the computational mining puzzle to solve, the safer and more stable is the institutional governance technology because it becomes more costly for a potentially dishonest miner to conduct an attack. Such an attack may adversely affect the perception of Bitcoin by its users reducing their trust and hence valuation of the cryptocurrency. If the reduction of the trust is large, it may cause a collapse in the economic value (price) of Bitcoin. As equation (11) implies, the amount of computer power for mining and hence the hash rate of the network would reduce, which might eventually lead to a collapse of Bitcoin blockchain. Thus, the security of Bitcoin blockchain is dependent on the size of mining reward received by miners which also determines the total computer power determined in equation (11).

## A.2.1 Blockchain security and attacks

The probability of a (successful) attack on blockchain is reflected in the underlying ledger's security, it is inversely related the blockchain's security budget. This probability is driven by the balance of computing power between an attacker and honest miners. As noted by BitcoinWiki (2019), "*Bitcoin's security model relies on no single coalition of miners controlling more than half the mining power*".[17] A miner who controls more than 50% of the total mining network computer power could exercise attack on blockchain that involves the addition of blocks that are somehow invalid or reverse previous accepted transactions ("majority attack"). Either the blocks contain outright fraudulent transactions, or they are added somewhere other than the end of the longest valid chain. A successful majority attacker could prevent (for the time that the attacker controls mining) confirmation of new transactions (e.g. by producing empty blocks) and reverse own transactions which potentially allows double-spending thus affecting all transactions that share the history with reversed transactions (BitcoinWiki 2019).

In our model, to control a majority power, equation (11) implies that an attacker must control more than 50% of the total mining network computer power, $An_t m_t$, where $A > 1$. If we assume that the attack takes the duration equal to $s$ block time, then the attacker's costs[18] are $sA(cn_t m_t + q_t n_t I_t) - (1 - \theta)qn_t m_t$ and the mining reward during the attack is $sp_t R_t$, where $\theta$ ($0 \leq \theta \leq 1$) represents the proportion of the mining technology, $m_t$, that can be recovered (reused, resoled, repurposed) after the attack.[19] The first term of the attacker's costs, $sA(c_t n_t m_t + qn_t I_t)$, includes energy and investment costs, while the second term, $(1 - \theta)qn_t m_t$, represents the loss related to the part of mining technology that cannot be recovered after the attack.

To des-incentivize and deter attacks on blockchain, the cost of an attack must be greater than the potential gain from an attack. Using the optimal condition (9), this implies the following incentive compatibility condition for blockchain against attacks:

(13) $$sAnm^*[(c_t + q_t\delta) - (1-\theta)q_t] \geq (1-\Delta)sE(p_t)R_t + V_A(\Delta)$$

---

[17] Although Bitcoin has not suffered from a majority attack, a number of Altcoins were subject to successful attacks in the past. For example, this was the case of the Bitcoin hard fork (Bitcoin Gold) in May 2018 (stealing $18 million worth of Bitcoin and other cryptos), Ethereum Classic (ETC) in January 2019 (double spending to over 200,000 ETC worth around $1.1 million), and Verge (XVG) was attacked several times in 2018 (with the biggest attack extracting about 35 million of XVG) (ViewNodes 2019).

[18] According to Crypto51 (2020), the theoretical cost of a 51% attack on Bitcoin is $ 413,908 per one hour.

[19] Note that if Bitcoin does not collapse after the attack, the mining equipment can be reused in continuing mining Bitcoin.

where $\Delta$ ($0 \leq \Delta \leq 1$) is the proportional decrease in the price of Bitcoin after the attack and $V_A$ is the expected payoff of the attack which is dependent on $\Delta$ and is equal to the sum of gains, $V_t(\Delta)$, obtained over the duration of attack $s$ with $V_A(\Delta) = \sum_s V_t(\Delta)$.[20] The payoff from the attack, $V_A$, can represent the gain from a Bitcoin double-spending or other type of gains (e.g. gain from a short sale of Bitcoin, gain in Bitcoin future markets from price fluctuation caused by the attack).

Using equation (10), the incentive compatibility condition (13) can be rewritten as:

(14) $$\left\{[E(p_t)R_t]^{\frac{\gamma}{1-\gamma}} - \frac{(1-\Delta)}{\beta}\right\} sE(p_t)R_t \geq \frac{V_A(\Delta)}{\beta}$$

where $\beta = An_t[(c_t+q\delta)-(1-\theta)q]\left[\frac{\gamma(n_t-1)}{n_t^2}\frac{1}{c_t+(\rho+\delta)q_t}\right]^{\frac{1}{1-\gamma}}$

Consider an attack where the only gain, $V_A$, is double spending. The attacker acquires $X$ amount of Bitcoins which (s)he double spends during the attack by exchanging them for the standard fiat currency. This implies that the gain from attack is $V_A(\Delta) = E(p_t)X - \Delta E(p_t)X$. After the attack, the attacker keeps the value of (double spent) $X$ Bitcoins in the standard fiat currency, $E(p_t)X$, but loses partially or fully (value of) Bitcoins acquired for the attack, $\Delta E(p_t)X$. If $\Delta$ is sufficiently small (i.e. Bitcoin does not collapse after the attack), then the system is vulnerable to the double-spending attack. However, if $\Delta = 1$ there is no gain from double-spending attack because the double spending attacker loses exactly as much value as (s)he gains from double spending. That is, $V_A(\Delta) = 0$ and equation (14) collapses to $[(\gamma(n_t-1)/n_t^2)(E(p_t)R_t/c_t + (\rho+\delta)q_t)]^{1/(1-\gamma)} = m_t \geq 0$. If $\Delta$ is sufficiently large, then the attack can sabotage the blockchain and lead to its complete collapse if $\Delta = 1$. In this case, the motivation of the attacker may be other than the gain (profit) from double spending (e.g. adversary power interested to damage the Bitcoin which could include a competing centralized intermediary, a competing cryptocurrency, or other entity) (Budish 2018).

In line with Abadi and Brunnermeier (2018); Budish (2018), equation (14) implies that the equilibrium block reward to miners must be sufficiently large relative to the one-off gain from the attack. Given that the gain from the attack, $V_A(\Delta)$, is unknown (e.g. in the case of the double spending attack, $X$ an thus $V_A(\Delta) = E(p_t)X - \Delta E(p_t)X$ could be large for $\Delta < 1$) and its value might be substantial, the equilibrium mining power needs to be larger than the one implied by equation (11) in order to deter an attack. This is induced by the fact that the payoff from the blockchain attack, $V_A$, does not affect the economic behavior (incentives) of honest miners in allocating their computer power for mining (i.e. $V_A$ does not enter in equation (11)).

---

[20] Note that in the steady state situation assumed in the incentive compatibility condition (13), implies that the discount rate $\rho$ cancels out with $V_t(\Delta) = V_l(\Delta)$ for $t \neq l$.